# $1/f^s$ noise from random R-C networks driven by white noise current, with low frequency characteristics changed by percolation

**Baruch Vainas[1]**

The Weizmann Institute of Science, Rehovot 76100, Israel

### Abstract

A model based on thermal fluctuations in conductors in random resistor-capacitor (R-C) networks has been shown to generate a $1/f^s$ noise with $0 \leq s \leq 1$, while in many real systems the noise exponent is in the $0 \leq s \leq 2$ range. The wider range of noise exponents is shown here to be generated using a model of random R-C networks driven by white noise current source, and having different compositions of resistors and capacitors. "C-rich" networks approach a brown noise, $1/f^2$ response, while "R-rich" networks approach a white noise, $1/f^0$ response. Random R-C networks containing equal numbers of resistors and capacitors generate the classic, $1/f$ pink noise. Thus, the compositionally unbiased R-C networks produce the ubiquitous $1/f$ noise. Below a limiting frequency, which is a function of the size of the network, the values of individual R and C elements, and their relative numbers in the network, the power-law frequency AC response of the network no longer holds, and the $1/f^s$ noise response, turns into either capacitor's ($1/f^2$), or resistor ($1/f^0$) response, depending on the nature of the persistent conductivity structures: series R-C pathways, or pure C and R percolation pathways.

## Introduction

Based on thermal fluctuations, the commonly used physical model for the $1/f$ noise in electrical conductors is modeled by the Johnson noise [1].
Using the noise analysis module of the Simetrix circuit simulation software it was shown that a $1/f^s$ noise with, $s < 1$, can be expected in random capacitors-resistors networks, with the exponent, s, which is related to the fraction of capacitors in the network [2]. Given that the fraction, s, is confined by definition within the [0,1] interval, this model does not include the s>1 case, which is a part of the full range for the $1/f^s$ noise which is characterised by, $0 \leq s \leq 2$.
In the following sections, a white noise driven circuit is suggested, that shows a $1/f^s$ noise with, $0 \leq s \leq 2$.

---

[1] On Sabbatical leave from Soreq Nuclear Research Center, Yavne, Israel

## The model and simulations

Rather than simulating thermal fluctuations in resistors, as in the previous work [2], a white noise generator is used in this work as a current source at the terminals of a random resistor-capacitor (R-C) network. Similar random networks were described in earlier publications [3-5] that have modelled the AC response of random R-C networks driven by AC current sources.

Prior to studying random R-C networks, consider the response of a 1-dimensional R-C transmission line to a white noise current input.

The power spectral density of the voltage, $S_v$ at the input of a R-C transmission line driven by a white noise current source can be written as,

$$S_v = i^2 R/2\pi f*C \qquad \text{eq. 1}$$

for a white noise driving current of flat spectral density, of magnitude i, R and C, the resistance and the capacitance of the individual elements, and f, frequency in $sec^{-1}$ units, as given in reference [6].

It should be noted that both the R-C transmission line, with an equal number of capacitors and resistors, and large random R-C networks that have an equal number of capacitors and resistors, were shown to have similar AC impedance characteristics, such as the constant phase angle (CPA) of current relative to voltage, which in both cases, of the R-C line, and the random, 1:1, R:C, composition R-C networks, give a phase angle of 45 deg over several decades of frequency. In both cases, the AC conductivity is characterized by a power law dependence on frequency, in the range of frequencies where the CPA holds. Thus, the AC complex conductance, **Y**, of a random R-C network is given by,

$$\mathbf{Y} = (j*2\pi f*C)^s (1/R)^{1-s} \qquad \text{eq. 2}$$

when expressed by the capacitance, C, and the resistance, R, of the individual elements in the network, and $j=\sqrt{-1}$, the imaginary unit [3-5].

For the case of the R-C line driven by a white noise current source, the resulting power density in eq. 1 is shown [7] to be equivalent to the mean square noise voltage,

$$V^2 = i^2 (abs(\mathbf{Z}))^2 \qquad \text{eq. 3}$$

To apply a similar expression to the complex impedance of a random R-C network, with the fraction, s, of capacitors C, and, 1-s, fraction of resistors, R, the conductivity, **Y** of eq. 2 is first converted to the complex impedance **Z** for the random R-C network given by, **Z** = 1/**Y**, and the absolute (abs) value of the complex impedance Z for this network, is then derived

abs(**Z**) = abs(1/**Y**) = abs((j*2πf*C)$^{-s}$ (1/R)$^{s-1}$),

for, s=1/2,
abs(**Z**) = abs((-j)$^{1/2}$ *(2πf*C)$^{-1/2}$ R$^{1/2}$)

$\qquad$ = (R/2πf*C)$^{1/2}$ $\qquad\qquad\qquad\qquad\qquad\qquad\qquad$ eq. 4

The mean square noise voltage drop at the terminals of the random R-C network, given a white noise driving current of magnitude, i, and s=1/2, is then,

V² = i²(abs(**Z**))² = i²R/2πf*C $\qquad\qquad\qquad\qquad\qquad\qquad$ eq. 5

of the same noise spectral density for, s=1/2, as, S$_v$ = const/f, in eq. 1, which is the pure 1/f noise.

**Experimental**

All simulations were done with Simetrix, Spice-like, analog circuit simulation software [3].
The white noise current source consists of a noise generator function of the voltage source in series with a large resistor (10 Mohms), to form an effective current source.
Noise generator parameters used for simulations:
Interval: 500nsec - time interval within which a single random value of voltage is generated as an output.
RMS: 20 V - root mean square random voltage amplitude of the noise generator output.
Start time: 0
End time: 200msec

The Simetrix simulation circuit is given in Fig. 1.
The output voltage is detected at the right hand side of the random network, at the VOUT terminal (as the voltage drop across the network), and used with the fast Fourier probe option to obtain a noise spectral density in the log V – log frequency plane, where, V, is the voltage magnitude of the FFT output. Results can also be plotted in the dB - log frequency plane. For a, 1V, the reference voltage (V$_0$), a 1mV amplitude in density plots corresponds to -60 dB power ratio according to definition
G$_{dB}$=20*log$_{10}$(V/V$_0$) for power ratio given in decibels from the amplitude, V, data. Given that the log of a ratio 10 between voltages corresponds to 20 dB power ratio, the slope of -1 in the log V – log frequency plane, corresponds to -20 dB in the dB - log frequency plane.

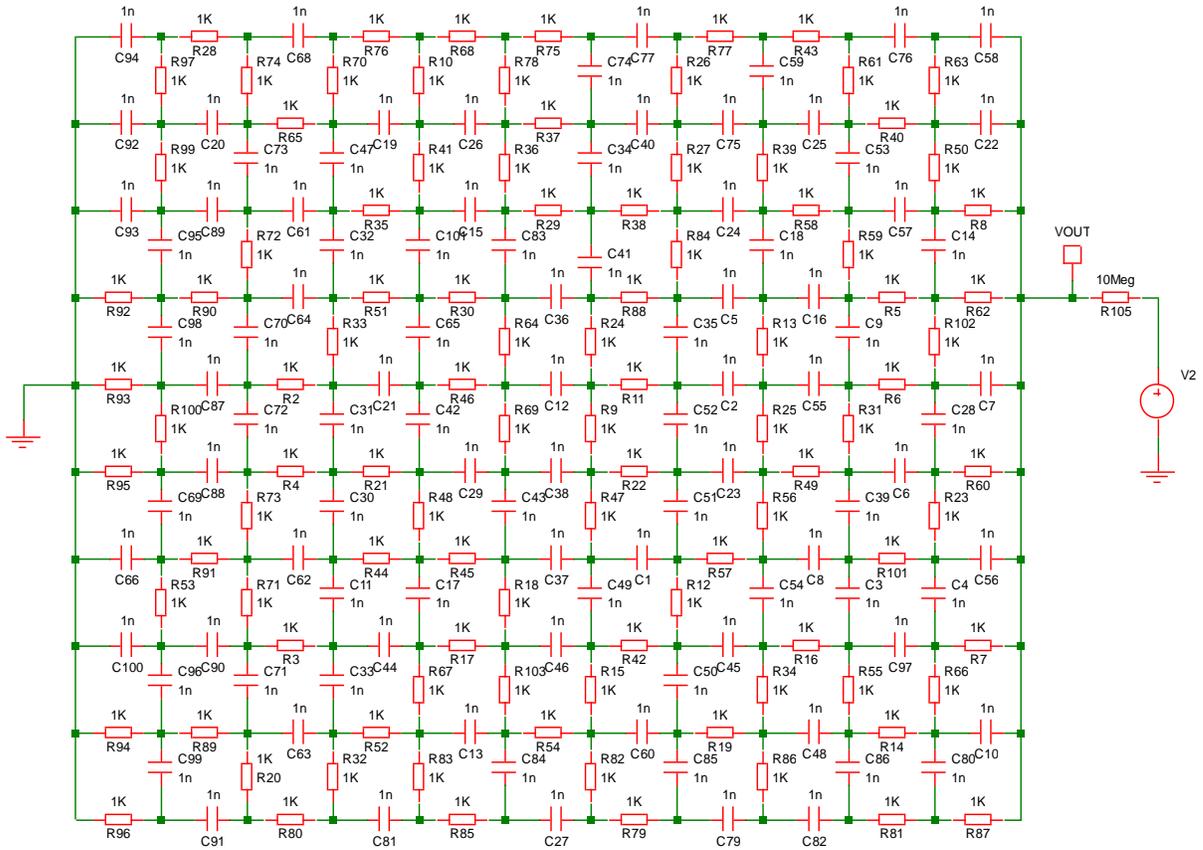

Fig. 1: Simetrix simulation of the response of a random R-C network to a white noise current source (noise voltage V2, in series with a large resistor R105), detected as the voltage drop VOUT across the network.

To begin, a pure resistor and a pure capacitor were used as a load to the white noise current source, instead of the random R-C network, to verify that a flat power spectrum for the resistor load is obtained. With a pure capacitor as a load, a 20dB/decade negative slope is expected, as the capacitor is an integrator for the white noise, resulting in a "brown" noise with, $S_v \sim f^{-2}$. Using the logarithmic form,
2log(V) = log(const) – 2log(f), the expected slope, in the log(V) vs log(f) plane, for the brown noise is -1.
As shown in Figs. 2-3, the expected power density characteristics are present. The two figures show the equivalent log voltage amplitude and dB representations respectively.

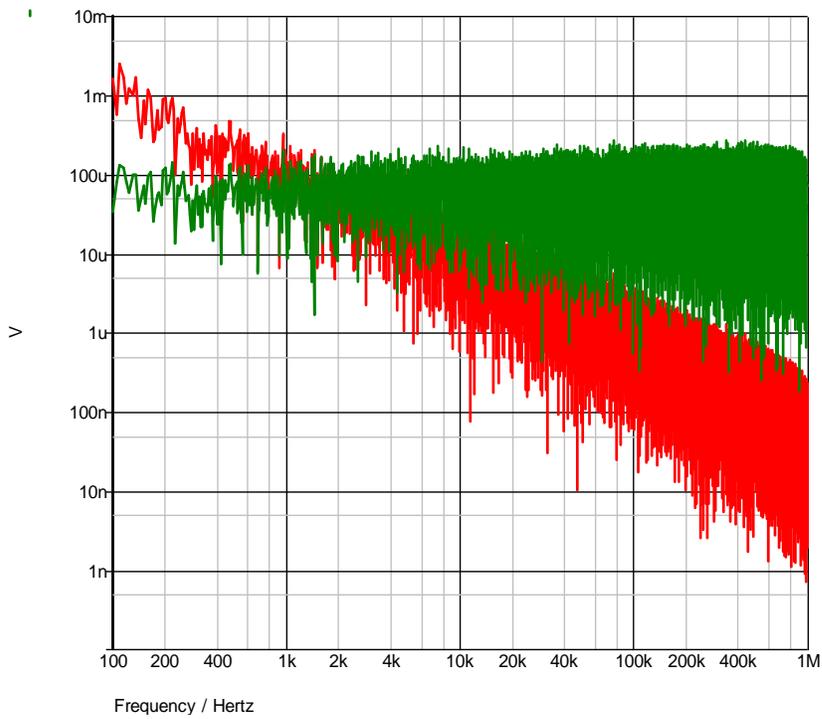

Fig. 2: Noise spectral density, log voltage amplitude vs. log frequency. Green trace - pure resistor, red trace - pure capacitor. Note the approximate -1 slope of the capacitor's response, vs the flat response of the pure resistor.

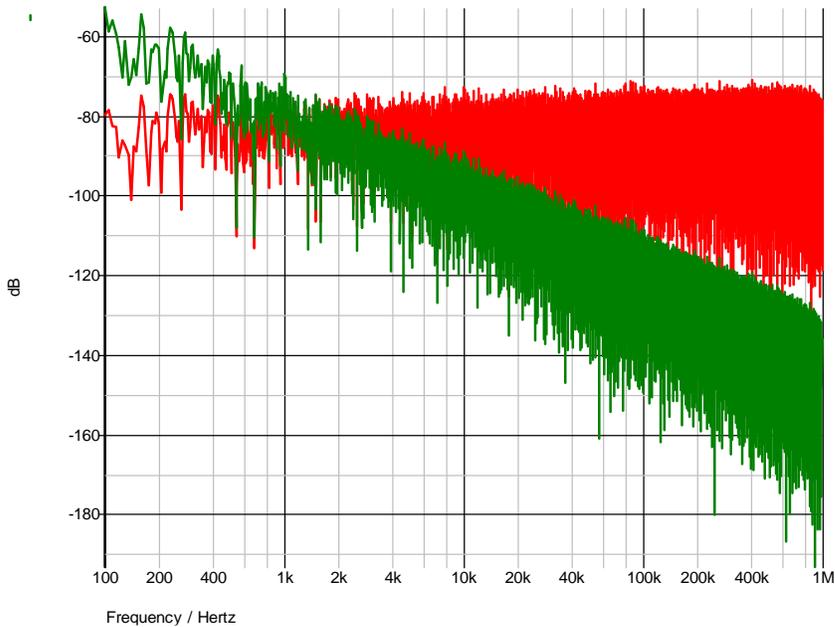

Fig. 3: Noise spectral density, dB units vs. log frequency. Green trace - pure capacitor,

red trace - pure resistor. Note the approximate -20dB/decade frequency slope of the capacitor's response, vs the flat response of the pure resistor.

The following graphs are given in a log voltage amplitude vs. log frequency representation only, with the response of the pure capacitor always included as a non-trivial reference, with the pure capacitor simulated simultaneously with the random R-C network.

In Fig. 4, a random network of Fig. 1 is analysed for its response to the white noise current generator. Given the, 1:1, C vs R, ratio in the network, a 1/f response is expected according to Eq. 5, which is represented by a slope -1/2 in the log(V) vs log(f) plane. The mean slope of the green trace in Fig. 4 is approximately -1/2 at frequencies above about 4 KHz, while below this frequency, the response is approximately flat, as for a pure resistor.
It should be noted that there is a single, R, percolation pathway in the network of Fig. 1.

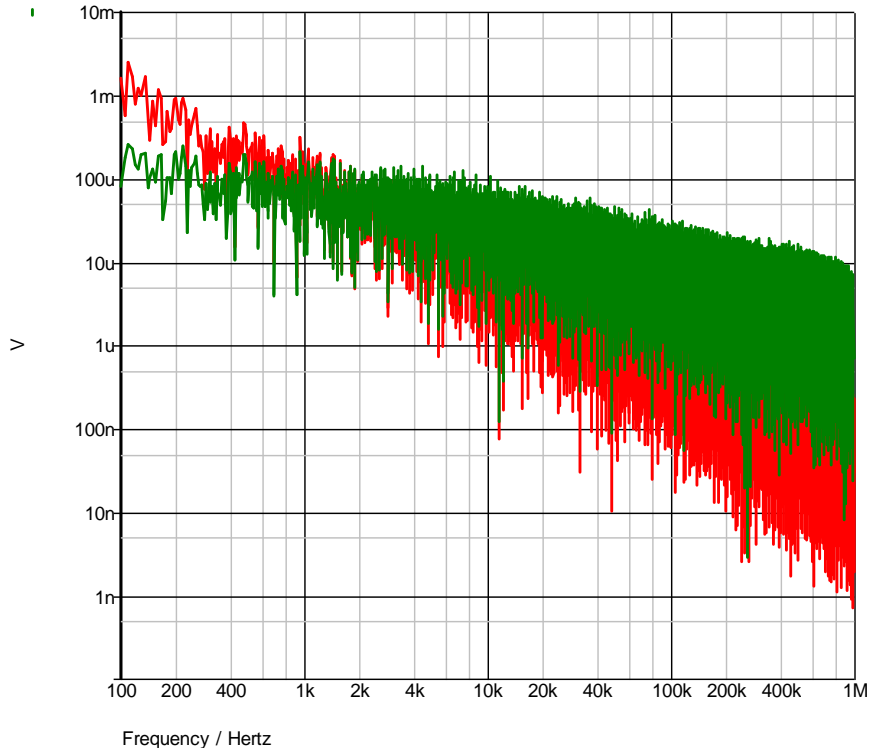

Fig. 4: Noise spectral density. Red trace – white noise applied to pure capacitor. Green trace – white noise applied to random R-C network. The network has a single conductive percolation pathway. Note an approximate slope -1/2 in the response of the random network at frequencies down to about 4 KHz. Below this frequency, the response is approximately flat.

To understand the flat response of the network below about 4 KHz, the conduction percolation pathway was eliminated by removing the resistor R20, on the conduction pathway in Fig. 1, and inserting a capacitor in its place. As can be seen in Fig. 5, the response of the network below approximately 4 KHz approaches the response of the pure capacitor, with the slope of -1. While the substitution of R20 with a capacitor did not create a capacitive percolation pathway, the $1/f^2$ – like noise spectral density is the expected result at the low frequency part of the spectrum, given that the response of series R-C network pathways are dominated by the high impedance of the capacitors at these frequencies.

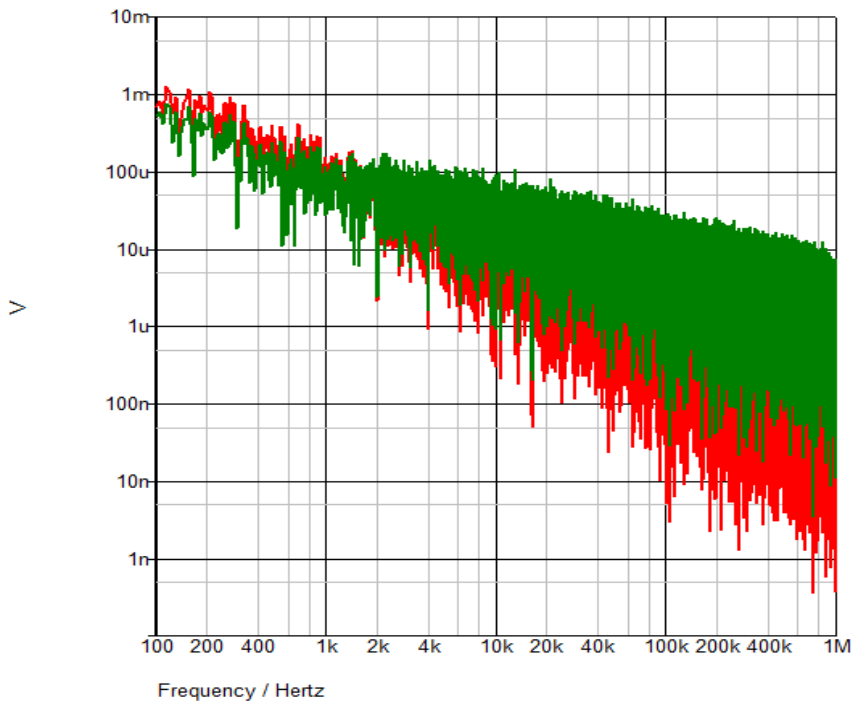

Fig. 5: Noise spectral density. Red trace – white noise applied to pure capacitor. Green trace – white noise applied to random R-C network. The network studied here does not have a conductive percolation pathway. Note an approximate slope -1/2 in the response of the random network at frequencies down to about 4 KHz. Below this frequency, the response approaches, -1, at the lower end of the frequency range, as for pure capacitor.

Random R-C networks have also been simulated for their response to white noise, with compositions differing from, 1:1, R:C.

For example, simulations of the C84:R116 network show an approximately flat spectral density below 10 KHZ, and given the percolation of conductive pathways of this R-rich network, a slope of slightly less negative than -0.5, above this frequency, is obtained as expected. This case is illustrated in Fig. 6, below.

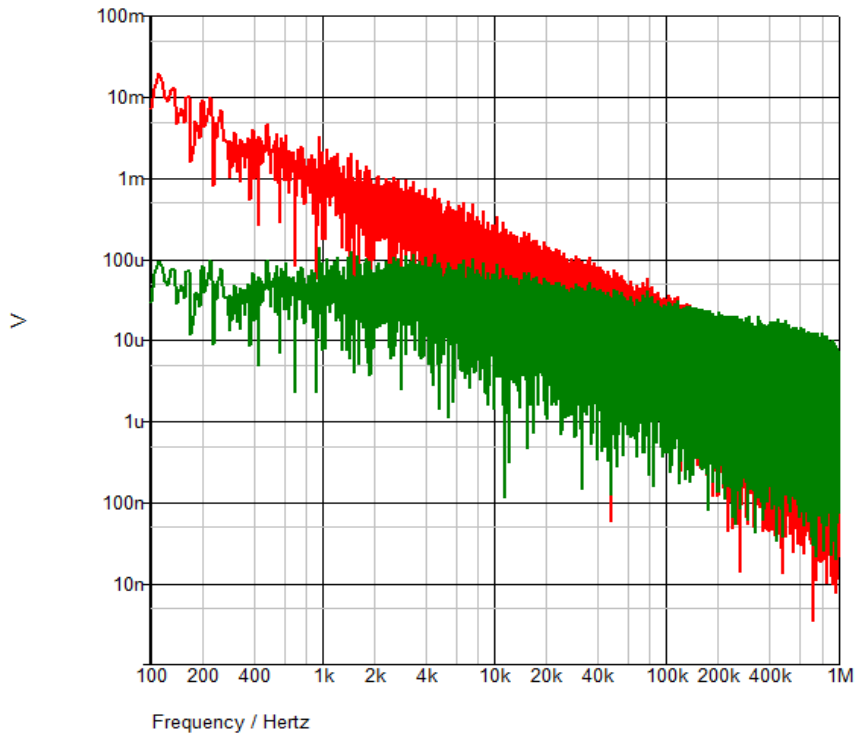

Fig. 6: White noise applied to pure capacitor - red trace, and C84:R116 random network –green trace. Note the flat spectrum of the R-rich network below about 10 KHz.

The 137C:63R (containing 137 capacitors and 63 resistors), capacitor-rich network had been simulated as well (Fig. 7), and this "C rich" network shows an intermediate, log amplitude – log frequency, characteristics, between that of the 1:1, slope -1/2 network and a pure capacitor, of slope -1. Given that this particular C-rich network tested has capacitors percolation pathways only, the low frequency characteristics (from about 10 KHz, and below) resemble a pure capacitor, as expected.

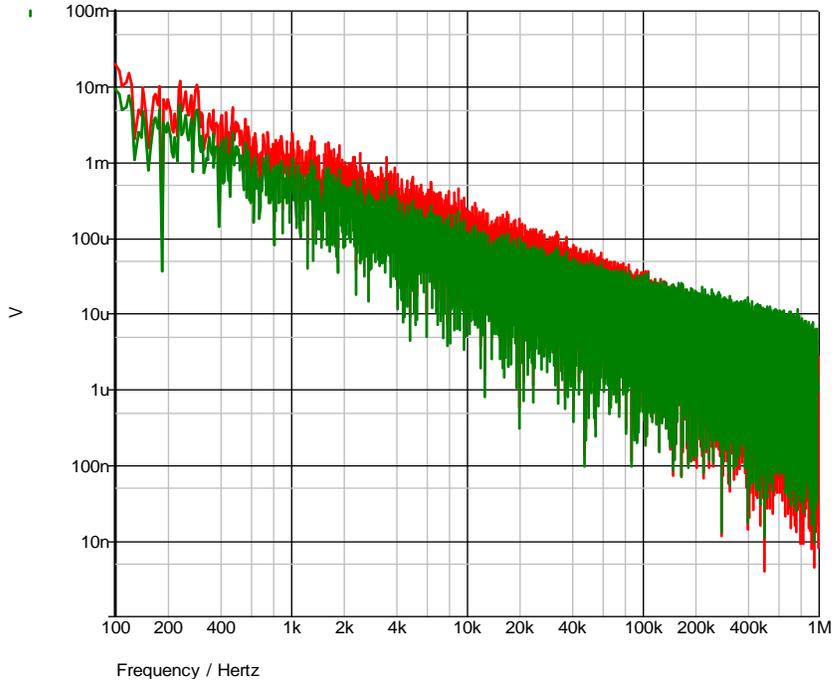

Fig. 7: White noise applied to pure capacitor - red trace, and 137C: 63R net random network –green trace. Note the -1 spectrum of the C-rich network below about 10 KHz.

It should be noted that the reduced span of the power law response, and the corresponding increase of the region of the low frequency flat, or -1 slope response, as shown in Figs. 6-7, in comparison with the, 1:1, R:C, case in fig. 4, was expected, as the validity range of the power-law impedance is reduced when the composition of the R-C network deviates from exactly dual, the 1:1, R:C composition.

**Conclusions**:

The power law of the frequency response of random resistor–capacitor networks can be easily utilized for the generation of the $1/f^s$ noise with the frequency components extending over wide ranges in the frequency domain. The widest range is obtained for networks containing equal numbers of capacitors and resistors for which, the power law AC impedance can extend to very low frequencies, without changing to a resistor, or a capacitor response. The low limiting frequency, at which the -1/2 power law response ends, in 1:1, R:C random networks, is determined [8] by the values of the individual capacitors (C), resistors (R), and the total number of impedance elements, N:

$f_{low} = 1/NCR$ eq. 6

Given the limited size of the network in this work, N = 200, the lower frequency limit of the power law impedance is given by eq. 6, as 5KHz (for, C=1nF, and R=1Kohm), which is close to the lower experimental limit for the 1/f noise of about 4Khz, as can be seen in Fig. 4. Obviously, extending the network would significantly lower the limiting frequency.

One can regard the 1/f generation method here as filtering a white noise by a power-law impedance of the random resistor-capacitor network, with the frequency range limits determined by the corresponding frequency limits of power law AC impedance of the network.

The power law region can extend without limits only for the, 1:1, R:C very large nets [8], generating the pure 1/f noise, while the power-law frequency response of non-1:1 nets is limited in range even for very large nets. Therefore, $1/f^s$, s≠1, noises are inherently limited in their frequency range, as can be observed in Figs. 6-7.

The ubiquity of the 1/f noise can therefore be partly, at least, explained by the extended range of frequencies it can be found.